\documentstyle[preprint,tighten,version2,aps]{revtex}
{\makeatletter%
\gdef\labeleqs#1{{%
\edef\@currentlabel{%
\ifappendixon\appletter\fi
\ifsecnumbers\ifnum\c@secnum>0
\arabic{secnum}.\fi\fi\arabic{equation}}%
\label{#1}%
}}%
}
\begin{document}
\draft
\preprint{IFUP-TH 16/94}
\begin{title}
Detecting Dual Superconductivity in the Ground State of Gauge Theory.
\end{title}
\author{L. Del Debbio, A. Di Giacomo and G. Paffuti}
\begin{instit}
Dipartimento di Fisica dell'Universit\`a and
I.N.F.N., I-56126 Pisa, Italy
\end{instit}
\begin{abstract}
We explicitly construct a monopole creation operator:
its vacuum expectation value is an order parameter for dual
superconductivity, in that, if different from zero, it signals a
spontaneous breaking of the $U(1)$ symmetry corresponding to monopole
charge conservation.

This operator is tested by numerical simulations in compact $U(1)$
gauge theory. Our construction provides a general recipe for
detection of the condensation of any topological soliton.
In particular our operator can be used to detect dual
superconductivity of the QCD vacuum.
\end{abstract}

\narrowtext
\section{Introduction}
A possible mechanism of colour confinement in Quantum Chromo-Dynamics
({\it QCD\/}) is dual superconductivity of the vacuum~\cite{tHO,Man,Nie,Rev}.

According to this scenario, the chromoelectric field is channelled
into Abrisokov flux tubes~\cite{Abri}, in the same way as the ordinary magnetic
field is in superconductors: the word {\it dual} indicates the
interchange of roles between electric and magnetic fields and charges.
The chromoelectric field mediating the force between coloured
particles is squeezed by Meissner effect into flux tubes of constant
energy per unit length, giving rise to the confining linear potential.
These flux tubes behave as strings~\cite{Nie,Nam}.
The existence of strings in hadronic physics is supported by phenomenology
\cite{Ven,Reb}. They have also been visualized by numerical simulations of QCD
on the lattice~\cite{dgmo,r10}. Some evidence in favour of dual
superconductivity of the vacuum has been produced by Montecarlo
simulations~\cite{r11}.

However a clear cut test of the mechanism is still missing. Such a
test would be the detection of monopole condensation in the ground
state, analogous to the condensation of Cooper pairs in the ground
state of an ordinary superconductor. Condensation implies that the
vacuum is a superposition of states with different charge, which, in
turn, is nothing but a spontaneous breaking of the $U(1)$ symmetry
related to charge conservation~\cite{Wein}. Such a breaking is
signaled by a non-vanishing vacuum expectation value
({\em vev \/}) of any operator carrying
non-trivial charge, as, e.g. the scalar field of the Landau-Ginzburg
model of superconductivity. That {\em vev} is called a disorder parameter
in the language of statistical mechanics.

In this paper, we give an explicit form of the creation operator of
monopoles, together with a lattice version of it, in view of using its
{\em vev \/} as a disorder parameter monitoring dual
superconductivity. The operator is checked in $U(1)$ compact pure
gauge theory on a lattice, where the behaviour of monopoles is
known~\cite{kmb,dGT}.

For {\it QCD\/}, the strategy is the following. The monopoles which are
expected to condense and generate superconductivity are Dirac
monopoles of a residual $U(1)$ symmetry, which survives after a
suitable gauge fixing, known as abelian projection~\cite{tHO}. An abelian
projection is defined as the gauge transformation which diagonalizes
any operator transforming in the adjoint representation of the gauge
group. The points where two eigenvalues of that operator coincide
correspond to singularities of the gauge transformation and have the
topology of world-lines of point monopoles~\cite{tHO,Schier}.
Such monopoles have been observed on the
lattice~\cite{r16,r17,ldd1,r19,r20}. Of course the location and the
number of  monopoles do depend on the choice of the operator used to define the
abelian projection: a possibility, advocated by 't~Hooft, is that
physics, i.e. monopole condensation, is independent of it. The
relevant abelian degrees of freedom
 can also be fixed by a somewhat different
procedure, known as {\em maximal abelian \/}
projection~\cite{Schier}.
Our operator will allow to investigate unambigously what abelian projection,
if any, defines the monopoles relevant to confinement.

In this paper,  we will only present the construction of the operator
and the way it works in lattice $U(1)$ compact gauge theory. For that theory,
there exists a construction of a disorder variable describing monopole
condensation~\cite{FM}, which is rigorous but based on a
particular form of the action (the Villain action). Our operator
coincides with the above one in the Villain case, but can be used with
different forms of the action. This is particularly important in order
to use it for non-abelian theories, where the effective $U(1)$ action
after abelian projection is not known. The study of monopole
condensation in {\it QCD\/}\ is in progress and will be presented elsewhere.

\section{The Monopole Creation Operator}

For the sake of definiteness, we shall consider here only the case of
a $U(1)$ monopole. Our procedure can easily be generalized to all kind
of solitons.

Let $B_{\mu}({\bf x}; {\bf y}) = (0, {\bf b}({\bf x}, {\bf y}))$ be
the classical field produced
in the location $\bf x$
by a Dirac monopole at
rest in  ${\bf y}$. We can make any choice for the gauge, e.g. by
putting the string along the positive $z$-axis. Then, defining
${\bf r} = {\bf x} - {\bf y}$ and $r =\|{\bf r}\|$, we have:
\begin{equation}
\label{eq:Dirac}
b_{i}({\bf x}, {\bf y}) = g\: \varepsilon_{3ij} \frac{r_{j}}{r(r-r_3)}
\end{equation}
where $g$ is the charge of the monopole, satisfying the Dirac
quantization condition:
\begin{equation}
e\, g = \frac{n}{2}, \; n \mbox{ integer}\label{defcarica}
\end{equation}

\noindent
If $\Pi_{i}({\bf x}, t)$ is the conjugate momentum to $A_{i}({\bf x},
t)$, then the operator
\begin{equation}
\mu({\bf y}, t) = \exp\{i \int d^{3}{\bf x} \; b_{i}({\bf x}, {\bf y}) \,
\Pi_{i}({\bf x}, t)\}
\end{equation}
creates a monopole at the location ${\bf y}$ and time $t$. This can be
immediately seen in the Schr\"odinger representation of the fields,
where we have:
\begin{equation}
\label{eq:trasl}
\mu({\bf y}, 0) |{\bf A}({\bf x}, 0)\rangle =
|{\bf A}({\bf x}, 0) + {\bf b}({\bf x}, {\bf y})\rangle
\end{equation}
Equation~(\ref{eq:trasl}) is a trivial consequence of the canonical
commutation relations between the fields and their conjugate momenta,
and is nothing but the field-theoretic equivalent of the familiar
statement that $e^{i\,pa}$ translates the
coordinate $q$ by $a$:
\begin{equation}
e^{i\,pa} |q\rangle = |q + a\rangle
\end{equation}
$\mu({\bf y}, t)$, applied to any field configuration, adds a monopole
to it.

\noindent
This can be restated in terms of commutation relations as:
\begin{eqnarray}
\left[A_{i}({\bf x}, t), \mu({\bf y}, t)\right] &=& b_{i}(x,y) \mu(\bf{y}, t)\\
\left[\Pi_{i}({\bf x}, t), \mu({\bf y}, t)\right] &=& 0
\end{eqnarray}
which also show that the electric field of the configuration is left
unchanged. It is worthwhile to notice that the specific choice of the
gauge for ${\bf b}$ is irrelevant: what really matters here is
topology, which is independent of it. Our operator $\mu$ is similar to
operators
introduced in the literature by different constructions, in various contexts
\cite{r22}.

\noindent
Now, if the ground state of the theory has a definite monopole number
$N$, then, under a magnetic $U(1)$ rotation, we have:
\begin{eqnarray}
U |0\rangle &=& e^{i \,\varphi \,N} |0\rangle \\
U \mu \, U^{\dag} &=& e^{i \,\varphi} \mu
\end{eqnarray}
Hence:
\begin{equation}
\langle 0| \mu |0\rangle = \langle 0| \mu |0\rangle e^{i \varphi}
\end{equation}
which implies:
\begin{equation}
\langle 0| \mu |0\rangle = 0
\end{equation}
Therefore if
\begin{equation}
\langle 0| \mu |0\rangle \neq 0
\end{equation}
then $|0\rangle$ is not $U(1)$ invariant and there is spontaneous
symmetry breaking of $U(1)$.
A translation by a static field ${\bf g}({\bf x})$ such that
$\mbox{curl }{\bf g}=0$ in this language corresponds to a pure gauge
transformation:
\begin{equation}
\gamma(t) = \exp \left\{ i \int d^{3}{\bf y} \; g_{i}({\bf x}) \,
\Pi_{i}({\bf x}, t)\right\}
\end{equation}

\begin{equation}
\gamma(t) | A_{i}({\bf x}, t) \rangle = | A_{i}({\bf x}, t) +
g_{i}({\bf x}) \rangle
\end{equation}
It is clear that $\gamma$ does not change ${\bf E}$ and, since ${\bf
H} = \mbox{curl } {\bf A}$, it does not change ${\bf H}$ either.
Therefore, by gauge-invariance,
\begin{equation}
\langle 0 | \gamma(t) | 0 \rangle = \langle 0 | 0 \rangle = 1
\end{equation}

\noindent
Performing the Wick rotation to Euclidean space, we obtain:
\begin{equation}
\mu_{E}({\bf x}, t) = \exp\left\{- g\;\int d^{3}{\bf y} \; b_{i}({\bf x},
{\bf y}) \, \Pi_{i}({\bf x}, x_4)\right\}
\end{equation}
{}From now on, we will be interested only in the Euclidean quantity and
drop for simplicity the $E$ subscript.

\noindent
For free photons, rescaling the fields by a factor $1/\sqrt{\beta}$
with $\beta = 1/e^2$,
\begin{equation}
\langle \mu \rangle = \frac{{\displaystyle \int} {\cal D}A \exp \left\{ -\beta
{\displaystyle \int}
\left[ \frac{1}{4} F_{\mu\nu} F_{\mu\nu} + F_{0i} b_{i} \right] \right\}}
{{\displaystyle \int} {\cal D}A \exp \left\{ -\beta {\displaystyle \int}
\left[ \frac{1}{4} F_{\mu\nu} F_{\mu\nu}\right] \right\}}
\end{equation}
The integral is Gaussian and can be directly computed giving:
\begin{equation}
\langle \mu \rangle = \exp \left\{ \frac{1}{2} \beta
\int \frac{d^{4}k}{(2\pi)^4} \langle F_{0i}(k)
F_{0j}(k) \rangle b_{i}({\bf k}) \; b_{j}({\bf k}) \right\}
\end{equation}
Since
\begin{equation}
\langle F_{0i}({\bf k}) F_{0j}({\bf k}) \rangle =
\frac{k_{4}^{2} \delta_{ij} + k_{i}k_{j}}{k^2}
\end{equation}
if ${\bf b}$ is such that $k_i b_i(k) = 0$, we have
\begin{equation}
\langle \mu \rangle = \exp \left\{ \frac{\beta}{2} \int
\frac{d^{4}k}{(2\pi)^4} |{\bf b}(k)|^2 -
\frac{\beta}{4} \int
\frac{d^{3}k}{(2\pi)^3} |{\bf k}| \, |{\bf b}(k)|^2 \right\}
\label{defmu}
\end{equation}
The same calculation for a pure gauge transformation yields:
\begin{equation}
\langle \gamma \rangle = \exp \left\{ \frac{\beta}{2} \int
\frac{d^{4}k}{(2\pi)^4} |{\bf g}(k)|^2 \right\}
\end{equation}
We realize  that the first term in the exponent
of eq.(\ref{defmu})
is a normalization
which can be subtracted by taking instead of $\langle \mu \rangle$ the
ratio:
\begin{equation}
\langle \bar\mu \rangle = \frac{\langle \mu \rangle}{\langle \gamma
\rangle}
\end{equation}
where $\langle \gamma \rangle$ is defined by means of any gauge
transformation ${\bf g}$, such that:
\begin{equation}
\label{eq:norm}
\int \frac{d^{4}k}{(2\pi)^4} |{\bf g}(k)|^2 =
\int \frac{d^{4}k}{(2\pi)^4} |{\bf b}(k)|^2
\end{equation}
Then
\begin{equation}
\label{eq:pert}
\langle \bar\mu \rangle = \exp \left\{
-\frac{\beta}{4} \int
\frac{d^{3}k}{(2\pi)^3} |{\bf k}| \, |{\bf b}(k)|^2 \right\}
\end{equation}
The integral in the exponent, once regularized at small distances,
tends to $+\infty$ as $V\to\infty$.
In the infinite volume limit $\langle \bar\mu \rangle=0$ as it should
be, since the perturbative vacuum has zero magnetic charge. The same
situation appears when computing the overlap of the Fock vacuum to the
Bogolubov rotated vacuum in a superconductor~\cite{r23}.

The same result is obtained if we work directly in the Schr\"odinger
representation in the Coulomb gauge. There the vacuum wave functional
is:
\begin{equation}
\Psi(\Omega) = \prod_{\bf k} C_{\bf k}
e^{- \frac{1}{2} |{\bf k}| \, |A_{\perp}({\bf k})|^2 }
\end{equation}
and
\begin{equation}
\langle \Omega | \mu | \Omega \rangle =
\exp \: \left\{- \frac{1}{4} \beta \int \frac{d^{3}k}{(2\pi)^3} |{\bf k}| \,
|{\bf b}(k)|^2 \right\}
\end{equation}

\section{Lattice Formulation}

A lattice version of the operator $\mu$ is obtained by  replacing
$e F_{0i}$ by the plaquette $P_{0i}$, or better, by its imaginary part:
\begin{equation}
e F_{0i} \to Im\,P_{0i}
\end{equation}
and discretizing the field $b_i$.

Then the disorder parameter becomes:
\begin{equation}
\langle \mu \rangle = \frac
{{\displaystyle \int} {\cal D}U \exp\left\{- \beta S \right\} \: \exp \left\{
-\frac{1}{2} \beta\sum_{n,i} b_{i}(n) Im\,P_{0i}(n) \right\}}
{{\displaystyle \int} {\cal D}U \exp\left\{- \beta S \right\}}
\end{equation}
or, if we want to cancel the unwanted normalization, we can divide by:
\begin{equation}
\langle \gamma \rangle = \frac
{{\displaystyle \int} {\cal D}U \exp\left\{- \beta S \right\} \: \exp \left\{
-\frac{1}{2} \beta\sum_{n,i} g_{i}(n) Im\,P_{0i}(n) \right\}}
{{\displaystyle \int} {\cal D}U \exp\left\{- \beta S \right\}}
\end{equation}
obtaining
\begin{equation}
\langle \bar\mu \rangle = \frac
{{\displaystyle \int} {\cal D}U \exp\left\{- \beta S \right\} \: \exp \left\{
-\frac{1}{2} \beta\sum_{n,i} b_{i}(n) Im\,P_{0i}(n) \right\}}
{{\displaystyle \int} {\cal D}U \exp\left\{- \beta S \right\} \: \exp \left\{
-\frac{1}{2} \beta\sum_{n,i} g_{i}(n) Im\,P_{0i}(n) \right\}}
\end{equation}
We stress once again  that any gauge function $g_{i}$ is acceptable,
provided the normalization condition (\ref{eq:norm}) is satisfied.

If we blindly compute $\langle \mu \rangle$ or  $\langle \bar\mu
\rangle$ by numerical simulations, a first technical difficulty
arises. We are faced with the usual problems encountered in computing
quantities like a partition function, which are exponentials of
extensive quantities, proportional to the number of degrees of
freedom. The distribution of the values is not Gaussian and the error
does not decrease by increasing statistics (see, e.g.~\cite{HHN},
where the same problem appears in a different context). To avoid that,
we will compute the quantity:
\begin{equation}
\rho = \frac{d}{d\beta} \log \langle \bar \mu \rangle =
\frac{d}{d\beta} \log \langle \mu \rangle -
\frac{d}{d\beta} \log \langle \gamma \rangle
\end{equation}
At $\beta=0$, $\langle \mu \rangle = \langle \gamma \rangle = 1$, and
therefore:
\begin{equation}
\log \langle \bar \mu \rangle = \int_{0}^{\beta} d\beta^{\prime}
\rho(\beta^{\prime})
\end{equation}
We easily get,  putting
\begin{eqnarray}
S_b &=& \left.\frac{1}{2} \sum b_{i}(n) Im\,P_{0i}(n)\right|_{n_0=0} \\
S_g &=& \left.\frac{1}{2} \sum g_{i}(n) Im\,P_{0i}\right|_{n_0=0}
\end{eqnarray}
the following expression for $\rho$:
\begin{equation}
\rho = \langle S + S_g \rangle_{S+S_g} -
\langle S + S_b \rangle_{S+S_b}
\end{equation}
which can be evaluated by numerical simulations. The subscript on the average
indicates the action defining the Feynman integral.

The two quantities on the {\em rhs} have the same strong coupling
expansion. Thus, the use of $\langle \bar\mu \rangle$ instead of $\langle
\mu \rangle$, besides producing a cancellation of the spurious
normalization of the Feynman path integral, can help in eliminating
the lattice artefacts produced by the discretization which can spoil the
continuum limit. This brings us to a second, more physical difficulty,
which is the continuum limit. In fact, while for {\it QCD\/}, which is
asymptotically free, we expect that, at sufficiently high $\beta$,
lattice artefacts should cancel, for a model like $U(1)$ this point is
not so clear in principle. In Ref.~\cite{FM} a proof is given of
monopole condensation in the confined phase of $U(1)$, defining a
disorder variable for the Villain action. We have checked that our
$\langle \mu \rangle$ operator exactly coincides with the one of
Ref.~\cite{FM}, when we use the Villain action: we expect that for
Wilson action the same will hold. We have then computed numerically
$\rho$.

For a {\em good} order parameter, we would expect $\rho$ to be zero,
or $\langle\bar\mu\rangle = 1$
below
the critical value of the coupling $\beta_c$ and then to show a large
negative peak around $\beta_c$, corresponding to a drop to zero of
$\langle \bar\mu \rangle$. At larger values of $\beta$, we have free photons
and Eq.~(\ref{eq:pert}) should hold.

Figure~1 shows the behaviour of $\rho$ for a $12^4$ lattice,
for a monopole in the center of the space lattice. In order
to be able to identify the signal as a genuine physical result (i.e.
not due to lattice artefacts), we have performed a number of checks:

\begin{itemize}
\item[1)] we have changed the form of $b_i$ by a gauge transformation to
get the Wu-Yang expression of the monopole potential. The result does
not change qualitatively.
\item[2)]  $\langle \gamma \rangle$ shows
practically no signal at $\beta_c$ within the errors, and does not change
appreciably by changes of $g_i$.
To test that we have computed $\langle \gamma\rangle$ for two different choices
of $g_i({\bf x})$,  both satisfying the condition Eq.(\ref{eq:norm}):
$\langle \gamma_1\rangle$ for ${\bf g}_1({\bf x}) = {\rm const.}$ and
$\langle \gamma_2\rangle$ for $g_2({\bf x})\propto {\bf x}/|{\bf x}|^2$.
In Fig.2 we display
\[ \rho_{gauge} = \frac{d}{d\beta}\ln\frac{\langle \gamma_2\rangle}
{\langle \gamma_1\rangle}\]
and $\rho = \frac{d}{d\beta}\ln\frac{\langle \mu\rangle}
{\langle \gamma_1\rangle}$ for a $6^4$ lattice.
$\rho_{gauge}$ shows no relevant signal at $\beta_c$.
\item[3)] We have measured the correlation function between a
monopole antimonopole pair at large time distance, in the same position in
space.
To do that we define
\[ C(d) = \langle\mu({\bf 0},0)\,\mu({\bf 0},d)\rangle\]
\[ S_{(b,\bar b)}(d) =
\left. \frac{1}{2}\sum b_i(n) Im P_{0i}(n)\right|_{n_0=0}
- \left. \frac{1}{2}\sum b_i(n) Im P_{0i}(n)\right|_{n_0=d}\]
and we measure
\[ \rho_{(b,\bar b)}(d) =
2 \langle S + S_g\rangle_{S+S_g} - \langle S + S_{(b,\bar b)}(d)\rangle_{
S + S_{(b,\bar b)}(d)}\]
Since
\[ \frac{d}{d\beta}\ln\frac{C(d)}{\langle\gamma\rangle^2} =
\rho_{(b,\bar b)}(d)\quad
\left. C(d)\right|_{\beta=0} = 1\quad
C(d) = \int_0^\beta \rho_{(b,\bar b)}(d) d\beta\]
By the cluster property we should have at large $d$ that
$C(d) \to \langle \mu({\bf 0},0)\rangle^2$, or
$\rho_{(b,\bar b)}(d) \to 2\rho$. Fig.3 shows that this expectation is indeed
verified.
\end{itemize}
We notice that the height of the negative peak of $\rho$ at $\beta_c$ increases
with volume [see Fig.1 and Fig.2]. The value of $\beta_c$ as defined by the
position of our peak is $\beta_c = 1.01(1)$ for a $6^4$ lattice and $\beta_c =
1.009(1)$ for a $12^4$ lattice.

We have taken monopole charge $n=4$ (Eq.(\ref{defcarica})) to get a good signal
with a relatively low statistics (tipically $10^4$ configurations per value of
$\beta$): smaller charges ($n=1,2$) give the same results but the signals are
smaller and more noisy. Our statistical errors are shown in the figures, when
they are larger than the symbols used.

\section{Conclusions}

The strong negative peak in $\rho$ at $\beta_c$ is a clear signal that
monopoles are condensed in the confined phase, while they are not in
the high-$\beta$ regime of the $U(1)$ model. The same signal can be
extracted from
$\langle \mu\,\bar\mu\rangle$ correlators.
We expect a similar signal in {\it QCD\/}\ at the deconfining phase
transition for the disorder variable creating $U(1)$ monopoles defined
by abelian projection. Work is in progress and a comprehensive
report will appear soon.

We underline that the technique used is exactly the same as for the
$U(1)$ and that it can be extended straightforwardly to different
systems in which the condensation of solitons can play a physical
role, like the 3-d $XY$ model, or the 4-d Georgi-Glashow model.
Work is in progress also on these systems.
\bigskip
\par
\noindent
{\bf Acknowledgements:} We thank M.Maggiore and S. Olejnik for
collaboration in the early stages of this work. We thank
M.~Mint\-chev for useful discussions.



\figure{$\rho$ versus $\beta$ on a $12^4$ lattice
\label{fig1}}

\figure{$\rho$ and $\rho_{gauge}$ versus $\beta$ on a $6^4$ lattice.
\label{fig2}}

\figure{$\rho_{b\bar b}(d)$ versus $\beta$ at distances $d=4,7,9$, compared to
$2\rho$.
\label{fig3}}

\end{document}